\documentstyle[prl,eqsecnum,aps]{revtex}

\begin{document}
\author{Jian-Qi Shen $^{1,}$$^{2}$ \footnote{Electronic address: jqshen@coer.zju.edu.cn}}
\address{$^{1}$  Centre for Optical
and Electromagnetic Research, State Key Laboratory of Modern
Optical Instrumentation, \\Zhejiang University,
Hangzhou SpringJade 310027, P.R. China\\
$^{2}$ Zhejiang Institute of Modern Physics and Department of
Physics, Zhejiang University, Hangzhou 310027, P.R. China}
\date{\today }
\title{Solutions of the Schr\"{o}dinger equation for the time-dependent linear potential\footnote{I think that this
paper will be a supplement to the recent comment [Phys. Rev. A
{\bf 68}, 016101 (2003)] of Bekkar {\it et al.} on Guedes's work
[Phys. Rev. A {\bf 63}, 034102 (2001)] and Guedes's reply to
Bekkar {\it et al.}'s comment. It will be submitted nowhere else
for publication, just uploaded at the e-print archives.}}
\maketitle

\begin{abstract}
By making use of the Lewis-Riesenfeld invariant theory, the
solution of the Schr\"{o}dinger equation for the time-dependent
linear potential corresponding to the quadratic-form
Lewis-Riesenfeld invariant $I_{\rm q}(t)$ is obtained in the
present paper. It is emphasized that in order to obtain the
general solutions of the time-dependent Schr\"{o}dinger equation,
one should first find the complete set of Lewis-Riesenfeld
invariants. For the present quantum system with a time-dependent
linear potential, the linear $I_{\rm l}(t)$ and quadratic $I_{\rm
q}(t)$ (where the latter $I_{\rm q}(t)$ cannot be written as the
squared of the former $I_{\rm l}(t)$, {\it i.e.}, the relation
$I_{\rm q}(t)= cI_{\rm l}^{2}(t)$ does not hold true always) will
form a complete set of Lewis-Riesenfeld invariants. It is also
shown that the solution obtained by Bekkar {\it et al.} more
recently is the one corresponding to the linear $I_{\rm l}(t)$,
one of the invariants that form the complete set. In addition, we
discuss some related topics regarding the comment [Phys. Rev. A
{\bf 68}, 016101 (2003)] of Bekkar {\it et al.} on Guedes's work
[Phys. Rev. A {\bf 63}, 034102 (2001)] and Guedes's corresponding
reply [Phys. Rev. A {\bf 68}, 016102 (2003)].
\end{abstract}
\section{Introduction}
Recently, Guedes used the Lewis-Riesenfeld invariant
formulation\cite{Lewis} and solved the one-dimensional
Schr\"{o}dinger equation with a time-dependent linear
potential\cite{Guedes}. More recently, Bekkar {\it et al.} pointed
out that\cite{Bekkar} the result obtained by Guedes is merely the
particular solution (that corresponds to the null eigenvalue of
the linear Lewis-Riesenfeld invariant) rather than a general one.
In the comment\cite{Bekkar}, Bekkar {\it et al.} stated that they
correctly used the invariant method\cite{Lewis} and gave the
general solutions of the time-dependent Schr\"{o}dinger equation
with a time-dependent linear potential\cite{Bekkar}. However, in
the present paper, I will show that although the solutions of
Bekkar {\it et al.} is more general than that of
Guedes\cite{Guedes}, what they finally achieved in their
comment\cite{Bekkar} is still {\it not} the {\it general}
solutions, either. On the contrary, I think that their result
\cite{Bekkar} also belongs to the particular one. The reason for
this may be as follows: according to the Lewis-Riesenfeld
invariant method\cite{Lewis}, the solutions of the time-dependent
Schr\"{o}dinger equation can be constructed in terms of the
eigenstates of the Lewis-Riesenfeld (L-R) invariants. It is known
that both the squared of a L-R invariant (denoted by $I(t)$) and
the product of two L-R invariants are also the invariants, which
agree with the Liouville-Von Neumann equation $\frac{\partial
}{\partial t}I(t)+\frac{1}{i}\left[ I(t),H(t)\right] =0$, and that
if $I_{a}$ and $I_{b}$ are the two L-R invariants of a certain
time-dependent quantum system and $|\psi(t)\rangle$ is the
solution of the time-dependent Schr\"{o}dinger equation
(corresponding to one of the invariants, say, $I_{a}$), then
$I_{b}|\psi(t)\rangle$ is another solution of this quantum system.
So, in an attempt to obtain the {\it general} solutions of a
time-dependent system, one should first analyze the complete set
of all L-R invariants of the system under consideration.
Historically, in order to obtain the complete set of invariants,
Gao {\it et al.} suggested the concept of basic invariants which
can generate the complete set of invariants\cite{Gao}, as stated
in Ref.\cite{Gao}, the basic invariants can be called invariant
generators. As far as Bekkar {\it et al.}'s result\cite{Bekkar} is
concerned, the obtained solutions are the ones corresponding only
to the linear invariant ({\it i.e.}, $I_{\rm l
}(t)=A(t)p+B(t)q+C(t)$) that is simply one of the L-R invariants,
which form a complete set. It is apparently seen that the
quadratic form, $I_{\rm
q}(t)=D(t)p^{2}+E(t)(pq+qp)+F(t)q^{2}+A'(t)p+B'(t)q+C'(t)$, is
also the one that can satisfy the Liouville-Von Neumann equation,
since it is readily verified that the Lie algebraic generators of
$I_{\rm q}(t)$ form a Lie algebra, which possesses the following
commutators
\begin{eqnarray}
\left[q^{2}, p^{2}\right]=2i(pq+qp),  \quad \left[pq+qp,
q^{2}\right]=-4iq^{2}, \quad
\left[pq+qp, p^{2}\right]=4ip^{2},    \nonumber \\
\left[q, p^{2}\right]=2ip,  \quad  \left[p, q^{2}\right]=-2iq,
\quad    \left[q, pq+qp\right]=2iq, \quad     \left[p,
pq+qp\right]=-2ip.                                 \label{algebra}
\end{eqnarray}
However, for the cubic-form invariant, it is easily seen that
there exists no such closed Lie algebra. This point holds true
also for the algebraic generators in any high-order L-R invariants
$I_{\rm l}^{n}$. So, it is concluded that for the driven
oscillator, only the linear $I_{\rm l}(t)$ and quadratic $I_{\rm
q}(t)$ will form a complete set of L-R invariants. Note that here
$I_{\rm q}(t)$ should not be the squared of $I_{\rm l}(t)$, {\it
i.e.}, $I_{\rm q}(t)\neq cI_{\rm l}^{2}(t)$, where $c$ is an
arbitrary c-number. It is emphasized here that Bekkar {\it et
al.}'s solution is the one constructed only in terms of the
eigenstates of the linear invariant $I_{\rm l}(t)$. Even though
only for the linear invariant $I_{\rm l}(t)$ Bekkar {\it et al.}'s
result\cite{Bekkar} can truly be viewed as the complete set of
solutions, it still cannot be considered general one of the
Schr\"{o}dinger equation, since the latter should contain those
corresponding to the quadratic invariant $I_{\rm q}(t)$. In brief,
Bekkar {\it et al.}'s solution and my solution, which will be
found in what follows, together constitute the complete set of
solutions of the Schr\"{o}dinger equation involving a
time-dependent linear potential.
\section{On the complete set of L-R invariants}
According to the L-R invariant theory\cite{Lewis}, if the
eigenstate of the linear invariant $I_{\rm l}(t)$ corresponding to
$\lambda_{n}$, {\it i.e.}, one of the eigenvalues of $I_{\rm
l}(t)$, is $|\lambda_{n}, t\rangle$, then the solution of the
Schr\"{o}dinger equation can be written in the form
\begin{equation}
|\Psi(t)\rangle_{\rm Schr}=\sum_{n}c_{n}\exp
\left[\frac{1}{i}\phi_{n}(t)\right]|\lambda_{n}, t\rangle
\label{2}
\end{equation}
with $c_{n}$'s and $\phi_{n}(t)$'s being the time-independent
coefficients and time-dependent phases\cite{Lewis,Gao},
respectively. This, therefore, means that the solutions of the
time-dependent Schr\"{o}dinger equation can be constructed in
terms of the complete set of eigenvector basis, $\{|\lambda_{n},
t\rangle\}$, of $I_{\rm l}(t)$. Moreover, one can readily verify
that the squared, $I_{\rm l}^{2}$, of the linear invariant is the
one satisfying the Liouville-Von Neumann equation, and that
$I_{\rm l}(t)|\Psi(t)\rangle_{\rm Schr}$ is also a solution (but
not another new general one) of the same time-dependent
Schr\"{o}dinger equation, since it is readily verified that
$I_{\rm l}(t)|\Psi(t)\rangle_{\rm Schr}$ can also be the linear
combination of the eigenstate basis set $\{|\lambda_{n},
t\rangle\}$ of $I_{\rm l}(t)$, {\it i.e.},
\begin{equation}
I_{\rm l}(t)|\Psi(t)\rangle_{\rm Schr}=\sum_{n}b_{n}\exp
\left[\frac{1}{i}\phi_{n}(t)\right]|\lambda_{n}, t\rangle,
\label{1}
\end{equation}
where the time-independent coefficients $b_{n}$'s are taken
$b_{n}=\lambda_{n}c_{n}$, which is obtained via the comparison of
the expression (\ref{1}) with (\ref{2}).

Thus, the above discussion shows that the linear invariant $I_{\rm
l}$ and its squared $I_{\rm l}^{2}$ have the same eigenstate basis
set and therefore $I_{\rm l}$ and $I_{\rm l}^{2}$ cannot form a
complete set of L-R invariants. In contrast, if for any c-number
$c$, the quadratic $I_{\rm q}$ cannot be written as the squared of
linear $I_{\rm l}$ with various integral constants $A_{0}$,
$B_{0}$ and $C_{0}$ (for the definition of $A_{0}$, $B_{0}$ and
$C_{0}$, see, for example, in Ref.\cite{Bekkar}), namely, the
relation $I_{\rm q}=cI_{\rm l}^{2}$ is always not true, then
$\{I_{\rm l}, I_{\rm q}\}$ is the complete set of L-R invariants,
which enables us to obtain the general solutions (complete set of
solutions) of the time-dependent Schr\"{o}dinger equation.

Perhaps someone will ask such question as, ``Does there really
exist such quadratic $I_{\rm q}$ that can always not be written in
the form  $cI_{\rm l}^{2}$?'' or ``Maybe any $I_{\rm q}$ that
satisfies the Liouville-Von Neumann equation can surely be written
as the squared of certain $I_{\rm l}$. Really?'' Now I will
discuss these questions. Consider a given quadratic invariant
$I_{\rm q}$ that is written $I_{\rm
q}(t)=D(t)p^{2}+E(t)(pq+qp)+F(t)q^{2}+A'(t)p+B'(t)q+C'(t)$ whose
time-dependent parameters are determined by the Liouville-Von
Neumann equation, and a certain linear invariant $I_{\rm l
}(t)=A(t)p+B(t)q+C(t)$, the squared of which is $I_{\rm
l}^{2}=A^{2}p^{2}+AB(pq+qp)+B^{2}q^{2}+2C(Ap+Bq+\frac{C}{2})$.
Since the functions $A$, $B$ and $C$ can also be determined by the
Liouville-Von Neumann equation, the only retained parts left to us
to determine is the integral constants $A_{0}$, $B_{0}$ and
$C_{0}$. Choose the appropriate integral constants in $A$, $B$ and
$C$, and let $I_{\rm q}$ be the squared of $I_{\rm l}$ (should
such case exist), and then we have
\begin{eqnarray}
D=cA^{2},  \quad       E=cAB,     \quad      F=cB^{2},
                \nonumber \\
   A'=2cAC,             \quad   B'=2cBC,       \quad
   C'=cC^{2}.                        \label{eqq6}
\end{eqnarray}
If a given $I_{\rm q}$ can really be written as the squared of
$I_{\rm l}$, the above six equations are just used to determine
the c-number $c$ and the suitable integral constants $A_{0}$,
$B_{0}$ and $C_{0}$ in the functions $A$, $B$ and $C$. It is seen
that there are only four numbers expected to be determined, and
that, in contrast, we have six equations. So, it is possible that
there exist potential parameters $c$ and $A_{0}$, $B_{0}$, $C_{0}$
which will not agree with Eqs.(\ref{eqq6}) always for a given
parameter set $\{D, E, F, A', B', C'\}$, or, for a given parameter
set $\{D, E, F, A', B', C'\}$ there are always no such parameters
$c$ and $A_{0}$, $B_{0}$, $C_{0}$ which satisfy Eqs.(\ref{eqq6}).
The existence of $I_{\rm q}$ that cannot be written as the squared
of any $I_{\rm l}$ is thus demonstrated.

So, in the above we indicate that such two invariants $I_{\rm l}$
and $I_{\rm q}$ (which are independent) form a complete set of L-R
invariants.
\section{Unitary transformation associated with L-R invariants}
Now I will solve the time-dependent Schr\"{o}dinger equation, of
which the time-dependent Hamiltonian\cite{Guedes} is given
\begin{equation}
H(t)=\frac{p^{2}}{2m}+f(t)q,
\end{equation}
by making use of the Lewis-Riesenfeld invariant
theory\cite{Lewis}. The time-dependent L-R invariant used here
takes the form
\begin{equation}
I_{\rm q}(t)=D(t)p^{2}+E(t)(pq+qp)+F(t)q^{2}+A(t)p+B(t)q+C(t).
\label{eq11}
\end{equation}
With the help of the Liouville-Von Neumann equation, one can
arrive at
\begin{eqnarray}
\dot{D}+\frac{2E}{m}=0,  \quad   \dot{E}+\frac{F}{m}=0, \quad
\dot{F}=0,
                 \nonumber \\
                 \dot{A}+\frac{B}{m}-2Df=0,    \quad      \dot{B}-2Ef=0,     \quad
                 \dot{C}-fA=0
\end{eqnarray}
with dot denoting the derivative with respect to time $t$. The
above six equations (referred to as the auxiliary
equations\cite{Gao}) can be used to determine all the
time-dependent parameters $A(t)$, $B(t)$, $C(t)$ and $D(t)$,
$E(t)$, $F(t)$.

In accordance with the L-R theory, solving the eigenstates of the
invariant (\ref{eq11}) will enable us to obtain the solutions of
the time-dependent Schr\"{o}dinger equation. But, unfortunately,
it is not easy for us to immediately solve the eigenvalue equation
of the time-dependent invariant (\ref{eq11}), for the invariant
(\ref{eq11}) involves the time-dependent parameters. So, in the
following we will use the invariant-related unitary transformation
formulation\cite{Gao}, under which the {\it time-dependent}
invariant in (\ref{eq11}) can be transformed into a {\it
time-independent} one $I_{V}$, and if the eigenstates of $I_{V}$
can be obtained conveniently, the eigenstates of $I_{\rm q}(t)$
can then be easily achieved.

Here we will employ two time-dependent unitary transformation
operators
\begin{equation}
V_{1}(t)=\exp [\eta(t)q+\beta(t)p],   \quad
V_{2}(t)=\exp [\alpha(t)p^{2}+\rho(t)q^{2}]          \label{eq21}
\end{equation}
to get a {\it time-independent} $I_{V}$.  The time-dependent
parameters $\eta$, $\beta$, $\alpha$ and $\rho$ in (\ref{eq21})
are purely imaginary functions, which will be determined in the
following subsections. Since the canonical variables (operators)
$q$ and $p$ form a non-semisimple Lie algebra, here the first step
is to transform $I_{\rm q}(t)$ into $I_{1}(t)$, {\it i.e.},
$I_{1}(t)=V_{1}^{\dagger}(t)I_{\rm q}(t)V_{1}(t)$, which no longer
involves the canonical variables $q$ and $p$, and the retained Lie
algebraic generators in $I_{1}(t)$ are only $p^{2}$, $pq+qp$,
$q^{2}$. Note that these three generators also form a Lie algebra
(see the commutators (\ref{algebra})) . The second step is to
obtain the time-independent $I_{V}$, which will be gained via the
calculation of $I_{V}=V_{2}^{\dagger}(t)I_{1}(t)V_{2}(t)$. In this
step, the obtained $I_{V}$ has no other generators (and
time-dependent c-numbers) than $p^{2}$ and $q^{2}$, namely,
$I_{V}$ may be written in the form $I_{V}=\varsigma(p^{2}+q^{2})$
with $\varsigma$ being a certain parameter independent of time. It
is well known that the eigenvalue equation of $I_{V}$ is of the
form $I_{V}|n, q\rangle=(2n+1)\varsigma|n, q\rangle$, where $|n,
q\rangle$ stands for the familiar harmonic-oscillator
wavefunction. Hence, the eigenstates of the time-dependent L-R
invariant $I_{\rm q}(t)$ in (\ref{eq11}) can be achieved and the
final result is $V_{1}(t)V_{2}(t)|n, q\rangle$ with the eigenvalue
being $(2n+1)\varsigma$.
\subsection{The calculation of $I_{1}(t)=V_{1}^{\dagger}(t)I_{\rm q}(t)V_{1}(t)$}
By the aid of the Glauber formula, one can arrive at
\begin{eqnarray}
I_{1}(t)&=&Dp^{2}+E(pq+qp)+Fq^{2}+[A+2i(E\beta-D\eta)]p+[B+2i(F\beta-E\eta)]q
\nonumber   \\
&+&C-[-i(B\beta-A\eta)+D\eta^{2}+F\beta^{2}-2E\beta\eta].
\end{eqnarray}
If the two relations
\begin{equation}
A+2i(E\beta-D\eta)=0,   \quad             B+2i(F\beta-E\eta)=0
\label{eq22}
\end{equation}
are satisfied, then we can obtain\footnote{In general, for the
case of three-generator Hamiltonian (the generators of which form
a non-semisimple algebra), the time-dependent c-number
$C(t)-[-i(B\beta-A\eta)+D\eta^{2}+F\beta^{2}-2E\beta\eta]$ in
$I_{1}(t)$ are vanishing. See, for example, in
Ref.\cite{Shenarxiv}, which is a special case of the present
problem.}
\begin{equation}
I_{1}(t)=D(t)p^{2}+E(t)(pq+qp)+F(t)q^{2}.
\end{equation}
It follows from (\ref{eq22}) that the time-dependent parameters in
the unitary transformation $V_{1}(t)$ are expressed by
\begin{equation}
\eta=\frac{EB-FA}{2i(E^{2}-DF)},    \quad
\beta=\frac{DB-EA}{2i(E^{2}-DF)}.
\end{equation}
\subsection{The calculation of $I_{V}=V_{2}^{\dagger}(t)I_{1}(t)V_{2}(t)$}
By using the Glauber formula, one can arrive at
\begin{equation}
I_{V}\equiv V^{\dagger}_{2}(t)I_{1}(t)V_{2}(t)={\mathcal
D}p^{2}+{\mathcal E}(pq+qp)+{\mathcal F}q^{2},
\end{equation}
where ${\mathcal D}$, ${\mathcal E}$ and ${\mathcal F}$ are of the
form
\begin{eqnarray}
{\mathcal
D}&=&D+\frac{4iE\alpha}{(16\rho\alpha)^{\frac{1}{2}}}\sinh
(16\rho\alpha)^{\frac{1}{2}}
+\frac{-8(F\alpha-D\rho)\alpha}{16\rho\alpha}\left[\cosh (16\rho\alpha)^{\frac{1}{2}}-1\right],                \nonumber \\
{\mathcal
E}&=&\frac{2i(F\alpha-D\rho)}{(16\rho\alpha)^{\frac{1}{2}}}\sinh
(16\rho\alpha)^{\frac{1}{2}}+E\cosh (16\rho\alpha)^{\frac{1}{2}},
\nonumber \\
{\mathcal
F}&=&F+\frac{-4iE\rho}{(16\rho\alpha)^{\frac{1}{2}}}\sinh
(16\rho\alpha)^{\frac{1}{2}}
+\frac{8(F\alpha-D\rho)\rho}{16\rho\alpha}\left[\cosh
(16\rho\alpha)^{\frac{1}{2}}-1\right],              \label{eqqq}
\end{eqnarray}
respectively. It follows that if the following two equations are
satisfied,
\begin{equation}
E=\zeta\sinh (16\rho\alpha)^{\frac{1}{2}},     \quad
\frac{2i(F\alpha-D\rho)}{(16\rho\alpha)^{\frac{1}{2}}}=-\zeta\cosh
(16\rho\alpha)^{\frac{1}{2}},                    \label{eq23}
\end{equation}
then the coefficients of $pq+qp$ in $I_{V}$ is vanishing. In order
that we can analyze the above equations (\ref{eq23}) conveniently,
the time-dependent parameters $\alpha$, $\rho$ (which are expected
to be determined) and $F$, $D$ are respectively parameterized to
be
\begin{equation}
\alpha=\frac{u\theta}{4},   \quad    \rho=\frac{v\theta}{4}, \quad
F=h\cosh (\sqrt{uv}\theta),     \quad    D=g\cosh
(\sqrt{uv}\theta).       \label{eq24}
\end{equation}
Substitution of the expressions (\ref{eq24}) into (\ref{eq23})
yields
\begin{equation}
E=\zeta \sinh (\sqrt{uv}\theta),     \quad
\frac{i(hu-gv)}{2\sqrt{uv}}=-\zeta,                 \label{eq25}
\end{equation}
which can determine $\zeta$ and $\theta$ (expressed in terms of
$E$, $h$, $g$ and $u$, $v$). It is noted that if the functions $u$
and $v$ are finally determined, then the time-dependent parameters
$\alpha$ and $\rho$ in the unitary transformation operator
$V_{2}(t)$ (\ref{eq21}) can be obtained.

In what follows we will determine $u$ and $v$ via setting
${\mathcal D}={\mathcal F}=\varsigma$ with $\varsigma$ being
constant ({\it i.e.}, time-independent). Insertion of (\ref{eq24})
into (\ref{eqqq}) will yield
\begin{equation}
D+\frac{hu-gv}{2v}[\cosh (\sqrt{uv}\theta)-1]=\varsigma,    \quad
F-\frac{hu-gv}{2u}[\cosh (\sqrt{uv}\theta)-1]=\varsigma.
\label{eq26}
\end{equation}
Eq.(\ref{eq26}) can determine the functions $u$ and $v$, although
the problem is very complicated. Here it should be noted that
$\theta$ which has been determined by (\ref{eq25}) is also the
function of $u$ and $v$. Thus, in principle, we can obtain the
time-dependent functions $\alpha$ and $\rho$ in the second unitary
transformation operator $V_{2}(t)=\exp
[\alpha(t)p^{2}+\rho(t)q^{2}]$.

Now under the unitary transformation $V_{1}(t)V_{2}(t)$ the
time-dependent invariant $I_{\rm q}(t)$ is changed into a
time-independent one, {\it i.e.},
\begin{equation}
I_{V}\equiv\left[V_{1}(t)V_{2}(t)\right]^{\dagger}I_{\rm
q}(t)\left[V_{1}(t)V_{2}(t)\right]=\varsigma(p^{2}+q^{2})
\end{equation}
whose eigenvalue is $(2n+1)\varsigma$ and the corresponding
eigenstate is $|n, q\rangle$ that is the familiar stationary
harmonic-oscillator wavefunction, and the eigenvalue equation of
the time-dependent invariant $I_{\rm q}(t)$ is thus given as
follows
\begin{equation}
 I_{\rm q}(t)V_{1}(t)V_{2}(t)|n, q\rangle=(2n+1)\varsigma V_{1}(t)V_{2}(t)|n, q\rangle.
\end{equation}
\subsection{The solutions of the time-dependent
Schr\"{o}dinger equation} According to the L-R invariant theory,
the particular solution $\left| n, t\right\rangle _{\rm Schr}$ of
the time-dependent Schr\"{o}dinger equation is different from the
eigenfunction of the invariant $I_{\rm q}(t)$ only by a phase
factor $\exp \left[\frac{1}{i}\phi _{n}(t)\right]$, the
time-dependent phase of which is written as
\begin{equation}
\phi _{n}(t)=\int_{0}^{t}\langle n, q|\left [
V_{1}(t')V_{2}(t')\right]^{\dagger}\left[H(t')-i\partial/\partial
t' \right]\left[V_{1}(t')V_{2}(t')\right]|n, q\rangle {\rm d}t'.
\end{equation}
This phase $\phi _{n}(t)$ can be calculated with the help of the
Glauber formula and the Baker-Campbell-Hausdorff
formula\cite{Wei,EPJD}.

The particular solution $\left| n, t\right\rangle _{\rm Schr}$ of
the time-dependent Schr\"{o}dinger equation corresponding to the
invariant eigenvalue $(2n+1)\varsigma$ is thus of the form
\begin{equation}
\left| n, t\right\rangle _{\rm Schr}=\exp \left[\frac{1}{i}\phi
_{n}(t)\right]V_{1}(t)V_{2}(t)|n, q\rangle.
\end{equation}
Hence the general solution of the Schr\"{o}dinger equation can be
written in the form
\begin{equation}
|\Psi(q, t)\rangle_{\rm Schr}=\sum_{n}c_{n}\left| n,
t\right\rangle _{\rm Schr},
\end{equation}
where the time-independent c-number $c_{n}$'s are determined by
the initial conditions, {\it i.e.}, $c_{n}= _{\rm Schr}\langle n,
t=0|\Psi(q, t=0)\rangle_{\rm Schr}$.
\\ \\

In the above we thus found the general solutions of the
Schr\"{o}dinger equation for the time-dependent linear potential,
which corresponds only to the quadratic-form invariant
(\ref{eq11}). It is concluded here that the solutions obtained
above does not form a complete set of solutions of this
time-dependent Schr\"{o}dinger equation, and that Bekkar {\it et
al.}'s solution and my solution presented here will constitute
together such complete set of solutions of the Schr\"{o}dinger
equation.
\section{Discussions and Conclusions}
(i) In the present paper we show that since Bekkar {\it et al.}'s
solution\cite{Bekkar} has not yet contain those corresponding to
the quadratic invariant, it is not the true general solution of
the Schr\"{o}dinger equation for the time-dependent linear
potential. Instead, it is the solution corresponding only to the
linear L-R invariant. The obtained solution here is the one that
corresponds to the invariant (\ref{eq11}), which is of the
quadratic form. Since the linear and quadratic invariants form a
complete set of L-R invariants, Bekkar {\it et al.}'s
solution\cite{Bekkar} and my solution presented here constitute
such complete set of solutions of the Schr\"{o}dinger equation
involving a time-dependent linear potential.
\\ \\

(ii) It is well known that in quantum optics there are three kinds
of photonic quantum states, {\it i.e.}, Fock state, coherent state
and squeezed state. From my point of view, the calculation of the
variations of creation and annihilation operators ($a^{\dagger}$,
$a$) of photons under the translation ({\it e.g.}, $V_{1}$ of
(\ref{eq21})) and squeezing transformation ({\it e.g.}, $V_{2}$ of
(\ref{eq21})) operators shows that the variations of $a^{\dagger}$
and $a$ are exactly analogous to that of space-time coordinate
variations under the translation, Lorentz rotation (boosts) and
dilatation (scale) transformation\cite{Fulton} and thus these
three quantum states (coherent, squeezed and Fock states)
correspond to the above three conformal transformations,
respectively. I think that this connection between them is of
physical interest and deserves further consideration.
\\ \\

(iii) Guedes recently stated that in order to obtain the general
solutions of the Schr\"{o}dinger equation one must follow the L-R
invariant theory {\it step by step}\cite{Guedes2}. I don't approve
of this point of view, however. Personally speaking, in fact, the
L-R method has only one step, namely, the particular solution of
the time-dependent Schr\"{o}dinger equation is different from the
eigenfunction of the invariant only by a time-dependent phase
factor. In Ref.\cite{Bekkar} and \cite{Shenarxiv}, although we
follow the L-R method step by step, what we obtained still cannot
be viewed as the general solutions of the Schr\"{o}dinger
equation. For this reason, I think that ``step by step'' is not
the essence of getting the general solutions of Schr\"{o}dinger
equation. Instead, the key point for the present subject is that
one should first find the complete set of all L-R invariants of
the time-dependent quantum systems under consideration. For some
systems in the Hamiltonian there may exist no such closed Lie
algebra as (\ref{algebra}), the complete set of exact solutions
can be found by working in a sub-Hilbert-space corresponding to a
particular eigenvalue of one of the invariants, namely, only in
the sub-algebra (quasi-algebra) corresponding to a particular
eigenvalue of this invariant will such time-dependent quantum
systems (which have no closed Lie algebra) be
solvable\cite{japan}. For the time-dependent quantum systems,
there are no other eigenvalue equations of Hamiltonian than that
of the L-R invariants with time-dependent eigenvalues. The
complete set of invariants, instead of the time-dependent
Hamiltonian, can describe completely the time-dependent quantum
systems. For this reason, it is essential to find the complete set
of invariants for the time-dependent Hamiltonian of a given
quantum system.
\\ \\

(iv) In the Ref.\cite{Guedes}, the author says that to the best of
his knowledge there was no publication reporting the solution of
the Schr\"{o}dinger equation for the system described by
$H(t)=\frac{p^{2}}{2m}+f(t)q $ without considering approximate
and/or numerical calculations. I think that this is, however, not
the true case. In the literature, at least in the early of 1990's,
Gao {\it et al.} had reported their investigation of the driven
generalized time-dependent harmonic oscillator which is described
by the following Hamiltonian
$H(t)=\frac{1}{2}[X(t)q^{2}+Y(t)(pq+qp)+Z(t)p^{2}]+F(t)q$\cite{Gao}.
It is believed that my solution presented here is only the special
case of what they obtained\cite{Gao}.
\\ \\

\textbf{Acknowledgements}  This project was supported partially by
the National Natural Science Foundation of China under the project
No. $90101024$.

\end{document}